# Single Alloy Nanoparticle X-Ray Imaging during a Catalytic Reaction

Young Yong Kim[1], Thomas F. Keller[1, 2], Tiago J. Goncalves[3], Manuel Abuin[1], Henning Runge[1], Luca Gelisio[1], Jerome Carnis[1], Vedran Vonk[1], Philipp N. Plessow[3], Ivan A. Vartanyants[1, 4], Andreas Stierle[1, 2]*

[1] Deutsches Elektronen-Synchrotron (DESY), D-22607 Hamburg, Germany.

[2] University of Hamburg, Physics Department, D-20355 Hamburg, Germany.

[3] Institute of Catalysis Research and Technology, Karlsruhe Institute of Technology, D-76344 Eggenstein-Leopoldshafen, Germany

[4] National Research Nuclear University MEPhI, Moscow, 115409 Russia

*corresponding author: Andreas Stierle andreas.stierle@desy.de

**Abstract:** The imaging of active nanoparticles represents a milestone in decoding heterogeneous catalysts' dynamics. We report the facet resolved, surface strain state of a single PtRh alloy nanoparticle on $SrTiO_3$ determined by coherent x-ray diffraction imaging under catalytic reaction conditions. Density functional theory calculations allow us to correlate the facet surface strain state to its reaction environment dependent chemical composition. We find that the initially Pt terminated nanoparticle surface gets Rh enriched under CO oxidation reaction conditions. The local composition is facet orientation dependent and the Rh enrichment is non-reversible under subsequent CO reduction. Tracking facet resolved strain and composition under operando conditions is crucial for a rational design of more efficient heterogeneous catalysts with tailored activity, selectivity and lifetime.

Heterogeneous catalysts play a decisive role for today's and future industrial scale energy production, conversion and storage, as well as for exhaust gas cleaning in environmental applications (*1*, *2*). Further on, they are involved in more than 80% of all chemical production processes (*3*). At the nanoscale, heterogeneous catalysts are composed of oxide supported, active nanoparticles of varying size, shape and composition (*4*). Under operating conditions, the catalyst is exposed to reactive gas mixtures at atmospheric or higher pressures and elevated temperatures rendering it a rather intricate, dynamical system. The structural complexity together with harsh reaction conditions often hampers an atomic scale understanding of catalytic reactions, which is at the heart of any rational design of future more efficient catalysts with tailored activity, selectivity and lifetime (*2*). To overcome these limitations, high resolution imaging techniques are required, compatible with realistic reaction conditions. Transmission electron microscopy (TEM) has made significant progress in the last years in the investigation of nanoparticles under ambient pressure catalytic reaction conditions (*5–7*), nevertheless representing a challenge for high resolution TEM in combination with elevated temperatures (*8*). On the other hand, coherent x-ray diffraction imaging (CXDI) has proven to be a very powerful method for the structural characterization of individual nanoscale objects with nanometer resolution (*9–12*). It allows investigating the three-dimensional (3D) crystalline electron density and strain field of a single nano object as demonstrated for experiments under oxidizing and reducing conditions (*13–17*). However, there is a lack of operando studies at the single nanoparticle level of a working catalyst.

Alloy based three-way catalysts for exhaust gas cleaning containing PtRh nanoparticles were reported to exhibit increased activity due to synergistic electronic effects (*18*). Their near surface composition is expected to vary for different gas surroundings, impacting on the activity (*8*, *18*, *19*). Here we demonstrate that the surface of an active, single PtRh alloy catalyst nanoparticle can be imaged using x-rays under operando catalytic flow conditions during the prototypical CO oxidation reaction. We resolve the nanoparticle size, shape and the facet orientation dependent evolution of the surface and bulk strain field while simultaneously probing the catalytic activity. A rigorous comparison with density functional theory (DFT) calculations allows us to correlate the facet surface strain with its compositional evolution and energetics.

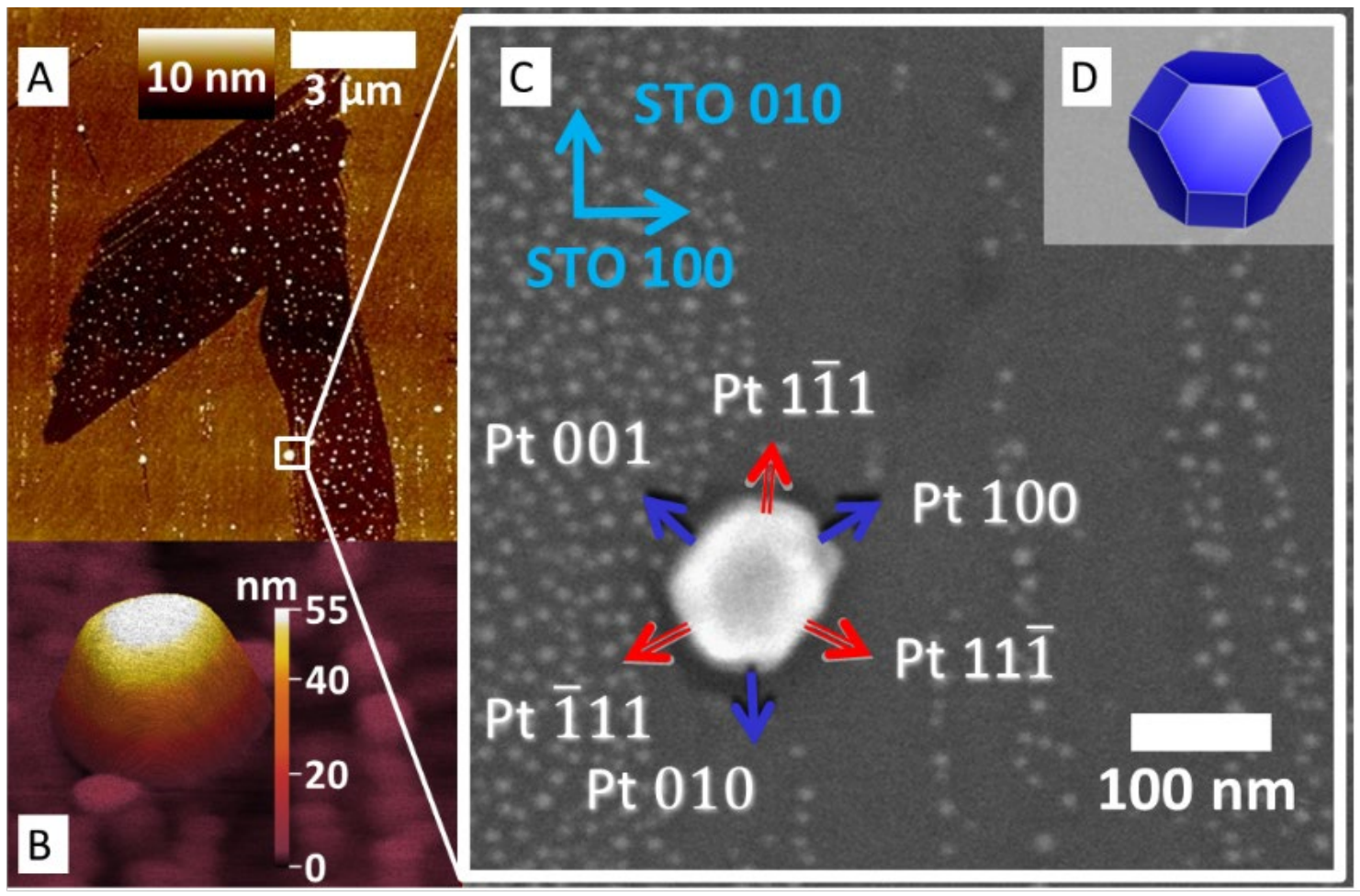

**Fig. 1. Sample architecture.** (**A**) Overview topographic AFM image containing the nanoparticle investigated during the CXDI experiment. **(B)** High resolution AFM image of the PtRh nanoparticle under investigation and smaller nanoparticles in the vicinity. **(C)** SEM zoom of the squared region marked in (A) together with nanoparticle facet and substrate directions. Blue arrows indicate <100> type facets (red arrows: <111> type facets). **(D)** Wulff-Kashiew construction of the nanoparticle equilibrium shape based on DFT surface energy calculations.

PtRh nanoparticles were prepared by co-deposition onto a $SrTiO_3$ (STO) (001) substrate, with a final annealing step at 1473 K to achieve equilibrium shape. Substrate defects, such as step edges, promote the growth of PtRh nanoparticles with diameters around 100 nm (see atomic force microscopy (AFM) image in Fig. 1A).

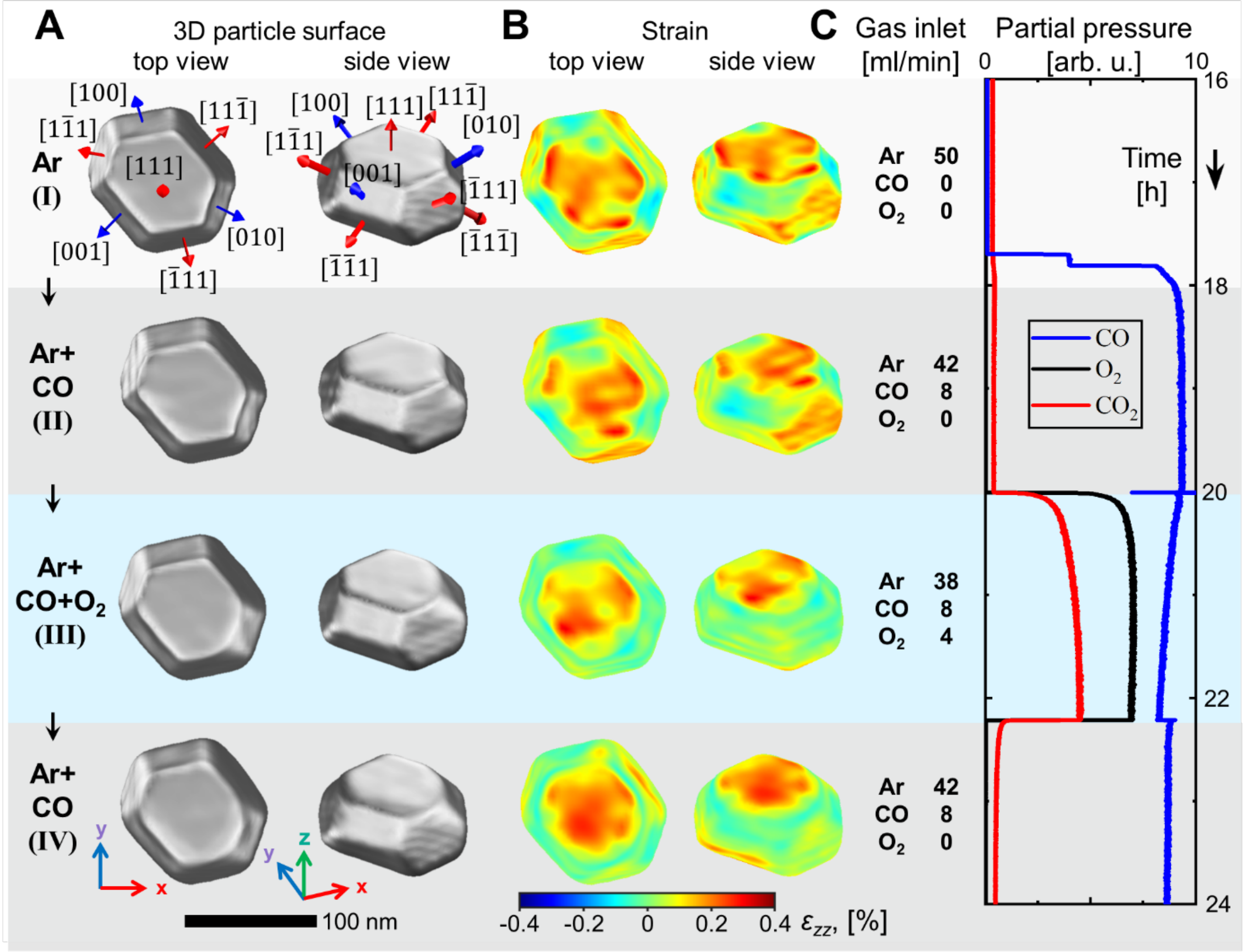

**Fig. 2. Nanoparticle shape and surface strain for different gas conditions.** (**A**) Top and side views of the reconstructed nanoparticle and (**B**) strain field $\epsilon_{zz}$ at the nanoparticle surface for gas conditions (I-IV). The position of the surface is defined as a cut at 55% of the reconstructed crystalline electron density from its maximum value. (**C**) Inlet gas composition and mass spectrometer signal during the experiment: CO (blue), $O_2$ (black) and $CO_2$ (red).

The CXDI catalysis experiment was performed at beamline ID01 at the European Synchrotron Radiation Facility (ESRF), Grenoble (France) using a nanofocused, monochromatic x-ray beam. The experiment was carried out at T=700 K and 50 mbar reactor pressure. The following gas conditions were applied to vary between reducing to active conditions: (I) pure Ar flow at 50 ml/min, (II) 8 ml/min CO and 42 ml/min Ar, (III) 8 ml/min CO, 4 ml/min $O_2$ and 38 ml/min Ar,

(IV) 8 ml/min CO and 42 ml/min Ar. Under stoichiometric reaction conditions (III) the ensemble activity of the catalyst was proven by a high $CO_2$ production, see Fig. 2C.

We selected one specific nanoparticle for in depth characterization, which we tracked by markers during the x-ray experiment. It is shown in the high-resolution AFM image in Fig. 1B and the scanning electron microscopy (SEM) image in Fig. 1C. The nanoparticle has a composition of $Pt_{60}Rh_{40}$ as determined by energy dispersive x-ray analysis, which is consistent with the lattice parameter determined by Bragg diffraction. The bulk of the nanoparticle exhibits a face centered cubic (fcc) structure, with statistical distribution of Pt and Rh atoms (*19*, *20*). The 111 Bragg reflection with scattering vector perpendicular to the top (111) surface of the nanoparticle was probed. The two-dimensional diffraction patterns were collected during rocking scans on an Eiger2M detector. They were converted to reciprocal space to obtain the 3D diffraction pattern from the investigated PtRh nanoparticle. The 3D nanoparticle crystalline electron density as well as the local strain component $\epsilon_{zz}$ in the direction of the scattering vector was determined by phase retrieval from the diffraction pattern averaged over the voxel size ($\approx 3\times3\times2$ nm$^3$).

In Fig. 2A the 3D nanoparticle shape is presented for the gas conditions I-IV, as obtained from the reconstructed electron density. The reconstructed nanoparticle is 55 nm high, 95 nm wide and 120 nm long, in very good agreement with SEM and AFM results (Fig. 1C). The facet angles can be precisely determined from cuts through the electron density, resulting in an exclusively low index termination of the nanoparticle by three <100> and eight <111> type facets, as indicated in Fig. 2A. The nanoparticle shape under pure Ar flow represents the as grown situation and is in good agreement with the DFT based Wulff-Kashiew construction (*21*) for a $Pt_{50}Rh_{50}$ nanoparticle on $TiO_2$ terminated STO (001), shown in Fig. 1D.

We focus now on the correlation of the PtRh nanoparticle facet surface strain evolution and their compositional change under catalytic reaction conditions. The reconstructed, facet resolved strain field $\varepsilon_{zz}$ at the nanoparticle surface is presented in Fig. 2B. Fig. 3 shows the facet resolved average surface strain and the strain distribution for the different reaction conditions. We make the interesting observation that under Ar flow conditions (I) the <111> type facets exhibit an average outward relaxation, whereas the <100> type facets are nearly strain free. All side facets exhibit a strain distribution width of ~0.1%. The top (111) facet strain distribution is much smaller than for the bottom ($\bar{1}\bar{1}\bar{1}$) facet, originating from the interfacial lattice mismatch. A similar situation is observed under Ar/CO flow conditions (II), in-line with the weak interaction of CO with the nanoparticle surface under the experimental conditions. The bulk lattice appears to be relaxed under Ar and Ar/CO conditions (the strain is close to zero), with the exception of a region with negative strain close to the top (111) facet in the middle of the nanoparticle, which may be attributed to a slight subsurface Rh enrichment (*22*).

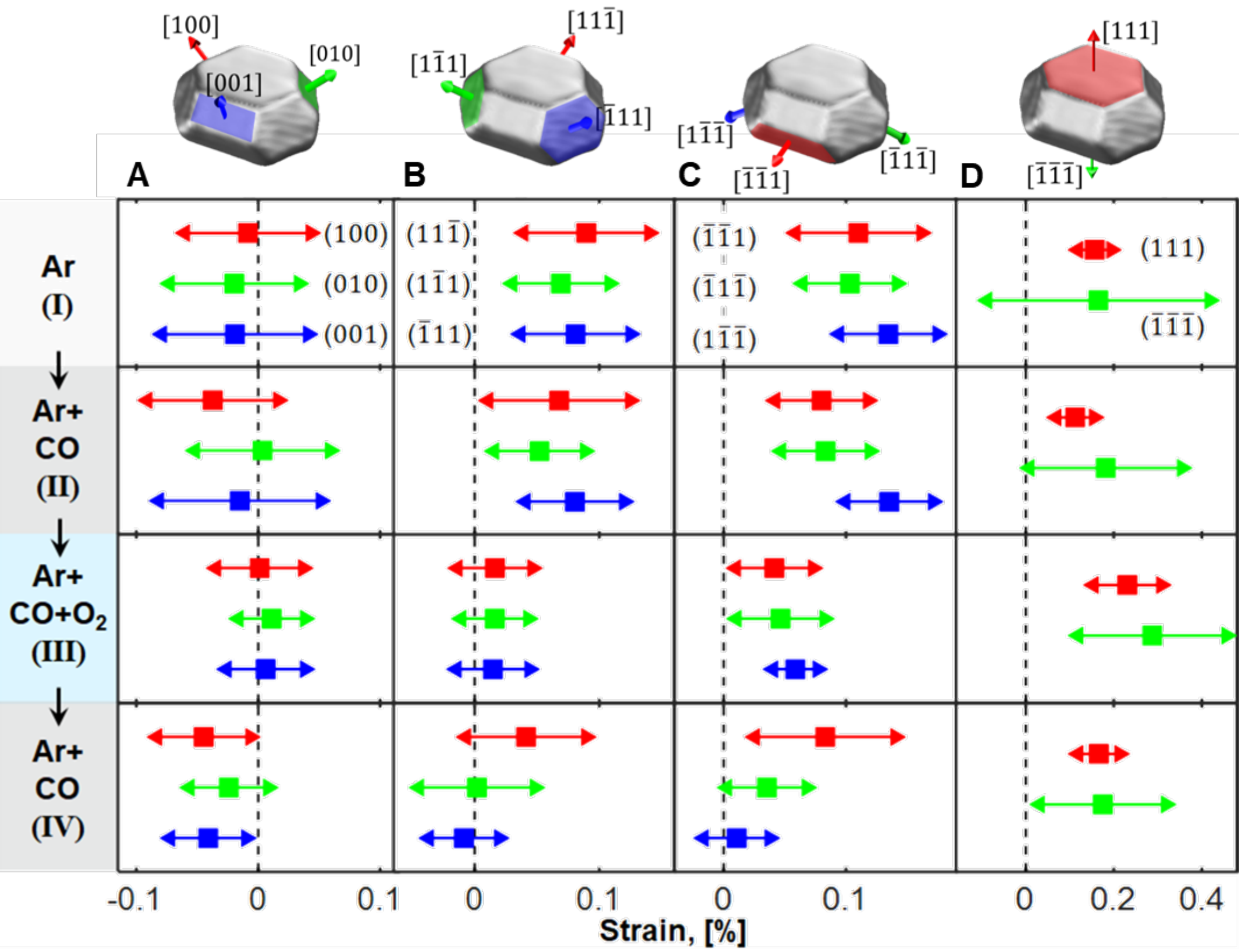

**Fig. 3. Facet resolved strain as a function of the reaction conditions.** (**A**): top <100> type side facets, (**B**) top <111> type side facets, (**C**) bottom <111> type side facets, (**D**) top (111) and bottom $(\bar{1}\bar{1}\bar{1})$ facet. Mean strain values: squares, standard deviation: lines with arrows. The colors of the facets in (A-D) correspond to the respective color of the experimental strain values in the plots below.

To elucidate the origin of the anisotropic facet surface strain pattern, we calculated a large library of structures by DFT using the PBE functional with dispersion corrections (*23*). We determined the most stable atomic structure and composition of the nanoparticle <100> and <111> type facets. Different terminations were studied by varying the composition of the outer two layers. Fig. 4 shows the most stable surface structures (Fig. 4A and 4B), the surface orientation, and adsorbate dependent layer resolved strain profiles for the most stable structures (Figs. 4C and 4D). To compare with the experimental data, the layer resolved strain values were averaged over the experimental voxel size (2 nm in z direction) and the theoretical strain component $\epsilon^t_{zz}$ in the direction of the scattering vector. To be able to compare with the experimental $\epsilon_{zz}$ values of the side facets, the theoretical $\epsilon_{zz}^{t}$ values have to be transformed to $\epsilon_{zz}^{t*} = \epsilon_{zz}^{t}\cos^2\alpha$, where $\alpha$ is the angle between the facet and the substrate ($\alpha = 54°$ for <100> type side facets and $\alpha = 70°$ for <111> type side facets). The theoretical results are robust against the choice of the slab size and the functional. Here we assume that the atoms mainly relax perpendicular to the facets, which is justified for larger facets. Fig. 4E and 4F show $\epsilon_{zz}^{t}$ trends for the clean and oxygen covered (100) and (111) surfaces: at higher Rh concentrations in the topmost layer, $\epsilon_{zz}^{t}$ gets more negative, in line with the 3.2% smaller atomic radius of Rh as compared to Pt. In the absence of adsorbates after deposition, in Ar-atmosphere or Ar/CO mixture (we expect no significant CO-coverage under our experimental conditions), a 100% Pt surface termination is energetically most favorable for both (100) and (111) type facets, in agreement with previous investigations (*24–26*). The theoretical segregation induced surface relaxation pattern given in Fig. 4 C and D exhibits distinct differences for the 111 and 100 surfaces: for the 111 surface, a more pronounced outward relaxation of the first Pt layer is observed. The calculations predict a positive value for $\epsilon_{zz}^{t}$ of +0.23% for the top (111) facet, $\epsilon_{zz}^{t*}$ value of +0.03% for the top <111> type side facets and a value of $\epsilon_{zz}^{t*}$ =+0.01% for the <100> type side facets. Overall, we find good agreement with the experimental observations (see Fig. 3).

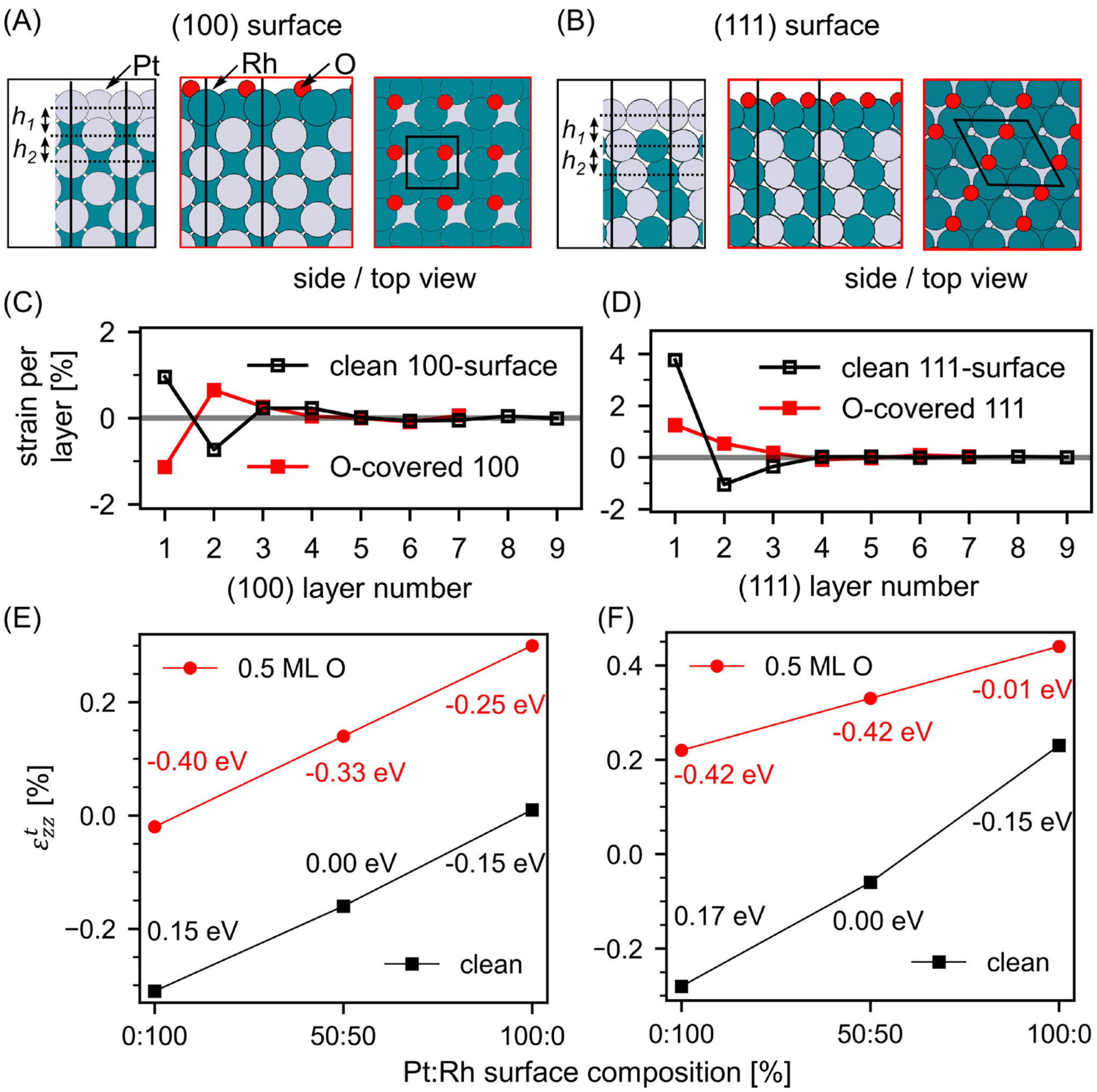

**Fig. 4. Surface composition and oxygen coverage dependent calculated surface strain.** (**A**) and **(B)** Side and top views of the clean and oxygen covered surfaces. (**C**) and (**D**) DFT (100) and (111) surface strain profiles for the clean and 0.5 ML oxygen covered (100) and (111) surface. (**E**) and (**F**) Trends in $\epsilon^t_{zz}$ for the (100) and the (111) surface as a function of the top layer Pt-Rh composition together with the surface energy difference per atom relative to the stoichiometric surface. Black squares: clean surface, red circles: 0.5 monolayer oxygen covered surface.

In our experiments, the bottom <111> type side facets exhibit more positive strain values as compared to the top <111> side facets, suggesting a stronger influence of the substrate or additional finite facet size induced strain effects. These effects were not included in the DFT calculations, which predict identical mean strain values for all <111> type side facets. The comparison between theoretical and experimental results clearly demonstrates that the CXDI surface strain pattern is directly connected to the surface composition induced relaxation state of the outmost surface layers of the nanoparticle. The comparison of the relaxation trends for <100> and <111> type facets provides evidence that under pure Ar and mixed Ar/CO flow, the whole nanoparticle surface is Pt terminated, which is also energetically most favorable under these conditions according to our calculations.

Next, we will discuss the evolution of the surface and bulk strain pattern under CO oxidation reaction condition (III). Importantly, the activity of the catalyst is evident from the emerging $CO_2$ mass spectrometry signal in Fig. 2C. The decreasing CO signal indicates self-activation during operation close to mass transfer limitations (*19*). Under reaction conditions, oxygen is dissociating on the Pt terminated nanoparticle surface *(27)*, and the higher affinity of oxygen to Rh as compared to Pt induces Rh surface segregation *(19, 28)*. We observe that under reaction conditions (third row in Fig. 2 and 3), strain is released over all nanoparticle side facets. This is most pronounced for the <111> type top side facets, with $\epsilon_{zz}$ values close to zero. The strain in the <111> type bottom side facets is less reduced. For the (111) top facet a strain reversal to more positive values is observed. The average interfacial strain slightly increased under these conditions. Another observation is the reduction of the strain distribution width under reaction conditions for the top <100> and <111> type facets to about half of the initial value, which may be related to a smoothening of the nanoparticle facets. We also observe that a rounding of the corners at the nanoparticle top takes place under reaction conditions, pointing to an autocatalytic self-tuning of the nanoparticle shape, providing more active step sites, especially at the edge between the (111) and the (010) side facet, see Fig. 2 *(5, 29)*. The strain in the nanoparticle bulk is observed to fully relax also in regions closer to the nanoparticle facets and the region of negative strain close to the top 111 facet in the middle of the nanoparticle disappeared, compatible with Rh surface segregation and concentration equilibration.

Figs. 4E and 4F show the theoretical $\epsilon_{zz}^{t}$ trends for 0.5 monolayer oxygen coverage as a function of the top most atomic layer composition. This represents the highest chemisorbed oxygen coverage to be expected under partially mass transfer limited reaction conditions. Also, for the oxygen adsorption structures, $\epsilon_{zz}^{t}$ gets more negative with increasing Rh concentration, but it is overall systematically shifted to more positive values. For both orientations, a 100% Rh surface composition is energetically more favorable in the presence of oxygen. O-Rh-O trilayer surface oxide formation is not compatible with our data (average strain >3%) *(19, 30)*.

The theoretical predictions rationalize our experimental observations: For the <100> type facets $\epsilon_{zz}$ of 0.01% is observed experimentally (compared to $\epsilon_{zz}^{t*}$ = -0.01% according to our calculations for the most stable structure with 100% Rh termination and oxygen on bridge site or a slightly less stable structure with $\epsilon_{zz}^{t*}$ = +0.02% for oxygen in hollow sites). In this case the oxygen adsorption induced increase in average strain is nearly fully compensated by the change in surface termination from 100% Pt to ≈100% Rh. Other, energetically less favorable surface compositions at 0.5 monolayer oxygen coverage give rise to theoretical $\epsilon_{zz}^{t*}$ values, that deviate by at least 0.04% from our experimental results.

For the oxygen covered (111) type facets, the calculations predict a 100% Rh and stoichiometric surface composition to be energetically degenerate, but more favorable than 100% Pt surface concentration, see Fig. 4F. The experimental $\epsilon_{zz}$ values of the <111> type side facets (0.02-0.05%) are compatible with a Rh surface composition of ~50%. The top 111 facet exhibits an average strain of 0.22%, pointing also to a 100% Rh composition.

In the next step of the reaction sequence, the oxygen flow was set to zero and the CO flow was kept constant. A break down in the $CO_2$ production rate was visible in the mass spectrometer signal in Figure 2C, condition (IV), accompanied by an increase in the CO signal due to a lifting of mass transfer limitations. Under CO rich conditions, all oxygen containing surface phases on the nanoparticle facets get reduced (*19*, *31*, *32*). The observed surface strain state of the nanoparticles, however, does not fully return to the previous situation under the same condition (II), see Fig. 3, bottom row. This points to an irreversible process that took place on the facet surfaces after complete oxygen reduction: we suggest that metallic Rh is still present at the surface, because of the reduced driving force for Rh subsurface segregation at 700 K (*22*, *24*). The <100> type facets exhibit a slightly negative experimental $\epsilon_{zz}$ of -0.03%, which is compatible with a $Pt_{75}Rh_{25}$ surface composition according to our calculations, see Fig. 4E ($\epsilon_{zz}^{t*} = -0.05\%$ for $Pt_{50}Rh_{50}$ to +0.01% for 100% Pt in the topmost layer). The strain state of the $(1\bar{1}1)$ and $(\bar{1}11)$ top side facets is also slightly reduced as compared to reaction conditions to values around zero. This is compatible with a surface composition slightly overstoichiometric in Pt, see Fig. 4F. The $(11\bar{1})$ top side facet and the $(\bar{1}\bar{1}1)$ bottom side facet on the opposite side of the nanoparticle, both oriented along the longer axis of the nanoparticle, exhibit a more positive strain state compatible with a 100% Pt surface concentration. The experimentally observed top surface 111 facet strain value is found to be 0.18%, pointing also to a 100% Pt surface composition.

We demonstrated that coherent x-ray diffraction imaging of a selected $Pt_{60}Rh_{40}$ alloy nanoparticle can be performed under realistic catalytic reaction conditions at elevated temperatures, retrieving the 3D nanoparticle shape and surface as well as bulk lattice strain state. Switching from reducing to active CO oxidation reaction conditions leads to a reduction of the surface strain state, which we resolve for individual <100> and <111> type facets. Through a rigorous comparison with DFT calculations, the facet strain state change can be attributed to gas environment induced nanoparticle surface compositional changes from a pure Pt termination under reducing conditions to a Rh rich termination under reaction conditions, varying for different facets. This process turns out to be non-reversible mainly for <111> type facets when switching back to reducing conditions. Such facet dependent, compositional heterogeneities may enable more effective reaction pathways and higher catalytic activity and selectivity. Our results open opportunities for surface sensitive single nanoparticle structural investigations of heterogeneous catalysts under industrial operando reaction conditions.